\documentclass[twocolumn,showpacs,preprintnumbers,prb]{revtex4}
\usepackage{amssymb}
\usepackage[dvips]{color,graphics,epsfig,rotating}
\usepackage{graphicx}
\usepackage{dcolumn}
\usepackage{bm}

\begin{document}

\title{The challenge of unravelling magnetic properties in LaFeAsO}
\author{I.I. Mazin}
\affiliation{Code 6693, Naval Research Laboratory, Washington, DC 20375}
\author{M.D. Johannes}
\affiliation{Code 6693, Naval Research Laboratory, Washington, DC 20375}
\author{L. Boeri}
\affiliation{Max-Plank-Institut f\"ur Festk\"orperforschung, Heisenbergstrasse 1, D-70569
Stuttgart, Germany}
\author{K. Koepernik}
\affiliation{IFW Dresden, P.O. Box 270116, D-01171 Dresden, Germany}
\author{D.J. Singh}
\affiliation{Materials Science and Technology Division, Oak Ridge National Laboratory,
Oak Ridge, Tennessee 37831-6114}
\date{\today }

\begin{abstract}
First principles calculations of magnetic and, to a lesser extent, electronic 
properties  of the novel LaFeAsO-based superconductors show
substantial apparent controversy, as opposed to most weakly or strongly
correlated materials. Not only do different reports
disagree about quantitative values, there is also a schism 
in terms of interpreting the basic physics  
of the magnetic interactions in this system. In this paper, we present a
systematic analysis using four different first principles  methods and show
that while there is an unusual sensitivity to computational details,
well-converged full-potential all-electron results are fully consistent
among themselves. What makes results so sensitive and the system so
different from simple local magnetic moments interacting via basic 
superexchange mechanisms is the itinerant character of the calculated 
magnetic ground state, where very soft magnetic moments and long-range
interactions are characterized by a particular structure in the reciprocal
(as opposed to real) space. Therefore, unravelling the magnetic
interactions in their full richness remains a challenging, but utterly
important task.

\end{abstract}

\maketitle

The disovery of high temperature superconductivity in LaFeAsO$_{1-x}$F$_{x}$
by Kamihara and co-workers \cite{kamihara} rising to critical temperatures $%
T_{c}$ over 50K with rare earth substitution, \cite{ren} has resulted in a
great deal of experimental \cite%
{ren,gchen0,sefat,gchen,cheng,xchen,gchen2,dong,zhu,cruz,mcguire} and theoretical
activity. \cite{singh,haule,mazin,xu,cao,ma,kuroki,dai,zhang,han,lee, boeri,
nekrasov,raghu,graser,yildirim,pickett} These studies
have shown that this set of Fe-based superconductors displays a remarkably
rich set of physical properties. Besides the high critical temperatures,
which include the highest known values of $T_{c}$ outside the cuprates,
there is evidence for several types of Fermi surface nesting \cite%
{mazin,kuroki} which raises the possibility of typically metallic
collective excitations, such as itinerant magnetization waves. At least
three different competing types of magnetic fluctuations have been predicted 
\cite{singh,haule,mazin,xu,cao,dong} and an ordered spin density wave was
first predicted\cite{mazin,dong} and subsequently observed experimentally in
the non-superconducting undoped parent, LaFeAsO\cite{zhu,cruz,mcguire}.
Currently, most researchers agree that superconductivity in this compound is
unconventional and likely related to magnetism \cite
{mazin,xu,cao,ma,kuroki,dai,zhang,han,lee,boeri,si,kivelson}.

The uncommon richness of the electronic structure of this compound has
already led to a situation in which different theoretical groups report
density functional theory (DFT) band structure calculations that emphasize
proximity to different magnetic states: weak ferromagnetism \cite%
{singh,haule} , checkerboard antiferromagnetism\cite{singh,mazin,cao}, an
antiferromagnetic stripe phase\cite{mazin,dong}. For the general reader, it
may look as if different DFT calculations disagree among themselves and
that, beyond a general initial consensus on the commonalities of the
different materials, electronic structure calculations differ in details
of the ground state and in the band structure near the Fermi level. The reason
for such inconsistencies is not just computational inaccuracy, as it 
may seem, but rather has a physical origin.
 Namely, as pointed out by a number of authors and discussed in
detail below, magnetism in this compound is very itinerant, making the
calculated magnetic energies and moments extremely sensitive to the
approximation used and to tiny details of the crystal structure. This
situation is relatively unusual, particularly in comparison to compounds with
localized magnetic moments, which are normally rather robust. Thus, there may
be a tendency to ascribe reported differences to an inherent inaccuracy of
electronic structure calculations, rather than to the peculiar dependencies
of the compound itself. Here we outline those dependencies and provide a
reference set of calculations in an attempt to establish a clear theoretical
picture within DFT.

There is a kernel of truth underlying the perception that DFT results are
disparate: for soft itinerant magnets one has to exercise extra caution in
the calculational parameters in order to get consistent and reliable
results. In particular, methods that imply restriction on the shape of the
charge density (such as Atomic Sphere Aproximation) and methods that involve
pseudization of the crystal potential can be trusted only after they have
been tested against all-electron, full-potential calculations. Furthermore,
in this particular compound the magnetic properties show an unusual
sensitivity to the internal As height and are also sensitive, unusual for
localized moments but typical for itinerant magnets, to the exchange
correlation potential.

Before investigating the magnetic properties and their dependencies in detail, it is important to note that this system is in 
proximity to a quantum critical point that results in strong spin fluctuations and to establish how such a system is best dealt 
with theoretically. It is well known (Refs. \onlinecite{Pd,NATO} and references therein) that DFT calculations, which are 
mean-field by nature, underestimate the effect of spin fluctuations that generally depress or suppress long-range magnetic order. 
It is also understood that gradient corrections (GGA) to the local density approximation generally increase the tendency to 
magnetism. As a rule of thumb, in systems with strong local Coulomb correlations, such as cuprates or 3d metal oxides, GGA better 
describes magnetic properties (an even further improved description is provided by the LDA+U method), while in itinerant magnets 
or near-magnets, LDA is closer to experiment in terms of magnetism. The classical examples are Pd metal and Sr$_{2}$RuO$_{4}$; 
both are ferromagnetic in GGA, but not (in agreement with the experiment) in LDA. On the other hand, GGA is usually better at 
predicting the crystal structure of transition metal compounds. Finally, given the same crystal structure, the bands themselves 
are practically indistinguishable whether calculated in LDA or GGA. With these facts in mind, previous experience in first 
principles DFT calculations suggests that the best approach to the LaFeAsO family may be to optimize the structure (if necessary) 
using GGA and to calculate magnetic properties using LDA (while LDA consistently overestimates the tendency to magnetism compared 
to the experiment, it always fares better than GGA). Either functional can be used to analyze bands and Fermi surfaces. If a 
quantitative value of the magnetic stabilization energy is less important than an accurate analysis of a \textit{trend }(say, 
with doping), GGA may be chosen for the reason that it amplifies the tendency to magnetism and results in larger numbers that are 
easier to compare with each other. What is important though is that as long one uses full-potential, all-electron methods (or 
well tested pseudopotentials) the results agree satisfactorily with each other, within the same density functional (LDA or GGA).  
It is worth noting that one can envision a situation where proper account for the energy associated with magnetic fluctuations 
may be more important than one-electron excitation spectrum (even when addressing the mechanism of superconductivity).  In such a 
case, spin-polarized GGA may give more insight than LDA (despite yielding an obviously exaggerated magnetic ground state); there 
is little previous experience to guide band theory procedure in this case.

Let us now discuss the reported results of magnetic calculations in more 
detail. Three different long range magnetic orders in the Fe plane have been 
so far considered: a ferromagnetic ordering that retains the crystal symmetry 
P4/nmm (\#129) with two formula units per cell, a checkerboard 
antiferromagnetism (P$\overline{4}$m2, \#115, two formulas), and an 
antiferromagnetism with alternating stripes along 100 (Pccm, \#49, four 
formula units). Singh and Du \cite{singh}, following the prescription above, 
presented LDA calculations for the undoped and 10\% doped materials. They 
found, using an in-house full-potential LAPW code, that the undoped material 
is barely unstable against weak ferromagnetism ($M\lesssim 0.1\mu _{B}$/Fe), 
but stable against checkerboard antiferromagnetism. Similarly, it was found 
that in LDA the ferromagnetic instability rapidly disappears with doping and 
at $x=$0.1 no longer exists. Mazin \textit{et al}\cite{mazin} later 
discovered that within LDA the doped material is unstable against the stripe 
AFM ordering.  It was shown subsequently shown \cite{dong} that this is the universal 
ground state \textit{ within DFT} at all dopings including zero.

\begin{table}[tbp]
\caption{Magnetic stabilization energies (vs. non-magnetic) and magnetic
moments for zero doping and hole ($x =$ 0.1) doping in the experimental and
optimized structures. Unfilled boxes correspond to unstable (vs. metastable)
configurations.}
\label{table_stab}
\begin{tabular}{ll|cc|cc}
\hline
optimized&   & \multicolumn{2}{c}{$x=0.0$} & \multicolumn{2}{|c}{$x=0.1$}   \\ 
\cline{3-6}
coordinates& & GGA & LDA                 & GGA & LDA   \\
 \hline
& FM & 0.1 & 0.0 & -- & --   \\ 
E$_{stab}$ & AFM (c) & 22 & -- & 10 & --   \\ 
(meV) & AFM (s) & 82 & 25 & 64 & 9   \\ \hline
& FM & 0.1 & 0.1 & -- & --   \\ 
$\mu$ & AFM (c) & 1.4 & -- & 1.3 & --   \\ 
($\mu$B/Fe) & AFM (s) & 1.8 & 1.2 & 1.8 & 1.0   \\ \hline \hline
experimental&  & \multicolumn{2}{c|}{$x=0.0$} & \multicolumn{2}{c}{
$x=0.1$}   \\ \cline{3-6}
coordinates&  & GGA & LDA & GGA & LDA   \\ \hline
& FM & 5.1 & 0.7 & 1.4 & 0.3   \\
E$_{stab}$ & AFM (c) & 87 & 26 &  85 & 28   \\
(meV) & AFM (s) & 180 & 84 & 171 & 75   \\ \hline
& FM & 0.3 & 0.2  & 0.2 &  0.1 \\
$\mu$ & AFM (c) & 1.8  & 1.5 & 1.8 &  1.5  \\
($\mu$B/Fe) & AFM (s) &2.2  & 1.8 &2.2  & 1.8   \\ \hline \hline
\end{tabular}
\end{table}

Soon after, Cao \textit{et al}, using pseudopotential methods, reported GGA
results for the checkerboard AFM ordering in the absence of doping. They
used two different techiques, VASP+PAW \cite{vasp} and PW-SCF ultrasoft
pseudopotential \cite{pwscf}. Neither of these methods is all-electron and
therefore depends on the selected pseudopotential (especially for Fe). Not
surprisingly, while qualitative conclusions agree with all-electron
calculations, quantitatively they differ both between themselves and with
the all-electron \textquotedblleft exact\textquotedblright\ result. Their
VASP calculations gave a very large magnetic stabilization energy of 84
meV/Fe, while PW SCF produced 14 meV/Fe.

\begin{figure}[tbp]
\includegraphics[height = .95\linewidth, angle=270]{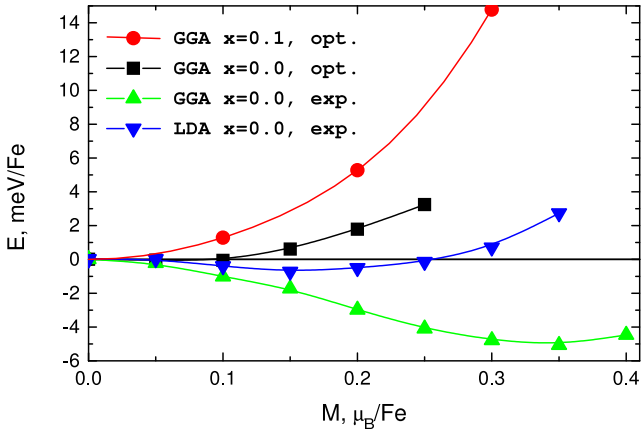}
\caption{(color online) Fixed spin moment calculations for LaFeAsO$_{1-x}$F$%
_x$.}
\label{FSM}
\end{figure}

Similarly, Dong \textit{et al} \cite{dong} reported pseudopotential calculations for the
stripe ordering proposed in Ref. \cite{mazin}, also in GGA, and found a
stabilization energy of 40 meV/Fe and magnetic moment of 1.5 $\mu _{B}$
(Dong \textit{et al} do not indicate what level of doping they were using,
but their plots are consistent with x=0.1).

An extensive study of both types of AFM ordering was reported by Yildirim \cite{yildirim} who performed both pseudopotential 
(which he discarded as 
unreliable) and all-electron calculations using LDA. Using fixed spin moment calculations, he found that only the stripe order is 
stable with respect to a non-magnetic state. Ma \textit{et al} \cite{ma}, using GGA and a pseudopotential approach, also found 
the stripe order 
to be the ground state, but also found a metastable checkerboard AFM state.

Finally, Yin \textit{et al}  \cite{pickett} performed GGA calculations and observed that
(a) the magnetic moment depends on the As position anomalously strongly and
(b) the As positions, optimized in GGA (as it is generally considered to be
the most accurate for the structural properties), differ from the
experimentally reported by as much as 0.1 \AA , an exceptionally large
discrepancy.

With this in mind, we have performed calculations using a variety of codes:
two entirely different all electron full-potential codes (WIEN2k\cite{wien2k}
and FPLO\cite{fplo}), as well as two different pseudopotential codes (VASP%
\cite{vasp} and PW SCF \cite{pwscf}). The two all-electron codes gave nearly
identical results, while the pseudopotential results differed somewhat,
though not drastically, both from all-electron and from each other to some
degree. We pursued the \textquotedblleft standard\textquotedblright route
discussed in the earlier section (using GGA for optimization and then
recalculating the magnetic properties using LDA). Optimization is performed
in the paramagnetic phase, since experimentally these materials are either
paramagnetic or have a drastically suppressed magnetic moment of 0.15-0.35 $%
\mu _{B}.$ The experimental lattice parameters, $a=$4.035 and $c$= 8.741 \AA %
\thinspace\ were used. We also compare our results to those in LDA (in the
same stucture) and to both LDA and GGA calculations in the experimental
structure.

Optimization leads to As and La positions that vary little with doping: from 
$z_{La}=0.1450,$ $z_{As}=0.6380$ at $x=0$ to $z_{La}=0.1490,$ $z_{As}=0.6340 
$ at $x=0.2.$ Reported experimental data are, respectively, 0.1415 and
0.6513 at zero doping, and practically the same at $x=0.1$ (ref.). As
pointed out by Yin \cite{pickett}, the discrepancy in the As position amounts to 0.1 
\AA , a nearly unheard-of an error for GGA calculations. In Table \ref%
{table_stab} and Fig. \ref{FSM} we show how the magnetic stabilization
energy varies as a function of doping for the optimized and experimental
structures. One sees that all GGA results are substantially more magnetic
(expectedly) as are the calculations in the experimental structure
(unexpectedly). A consistent set of numbers for calculations using
all-electron WIEN2k are given, with comparison numbers from other codes also
included. The two all-electron methods differ by about 10\% in terms of
magnetic energy; pseudopotential calculations are somewhat more off,
especially the PW SCF. Importantly, all of the them predict the same energy
sequence (this is however \textit{not} the case if a pseudopotential without 
$p$ states in the valence is chosen for Fe\cite{pp}). Finally, we compare the results
with published calculations and find the same trend: all electron
calculations agree reasonably well among themselves, while pseudopotential
ones, being sensitive to the choice of pseudopotential, scatter more.

Interestingly, apart from the structural parameters themselves LDA results
obtained using the GGA-optimized structure seem to correspond more closely
to what is physically observed than those obtained using the experimental
structure. General experience indicates that magnetism in metals near
quantum critical points can be entirely suppressed by fluctuations if the
LDA magnetic energy is on the order of 10-15 meV per atom, but it is rather
hard for spin fluctuations to entirely destroy magnetism that is stabilized
by any substantially larger energy. The stabilization energy for the stripe
antiferromagnetism in the experimental structure is 75 meV ($x=0.1),$ far
too much to expect full suppression of magnetic order from spin
fluctuations. In the theoretical structure, on the other hand, the same
energy is 9 meV, which is in precisely the range of energies that
accumulated knowledge of itinerant magnets indicates can be overcome by spin
fluctuations.

\begin{figure}
\includegraphics[height = 0.95\linewidth, angle = 270]{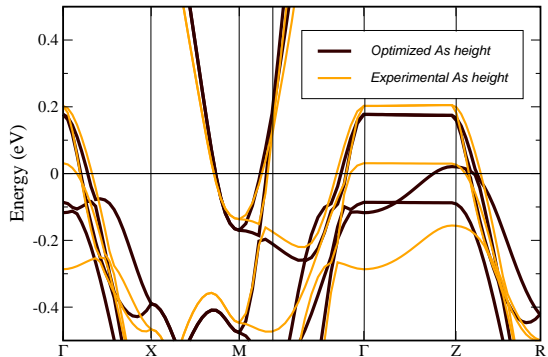}
\caption{(color online) A comparison of the band structure derived using experimental coordinates to that using optimized coordinates.  
The three-dimensional band that crosses the Fermi energy in the optimized structure is extraordinarily sensitive to As height, 
dropping below E$_F$ when it is raised to the experimental position.  It is replaced by a two-dimensional band.}
\label{oeband}
\end{figure}

The Fermi surfaces corresponding to the GGA-optimized structure in
comparison to those of the experimental structure further support use of the
optimized structure. An empirical observation can be made that the border
between the SDW antiferromagnetism and superconductivity in LaFeAsO$_{1-x}$F$%
_{x}$ is roughly at the same concentration ($x\sim 0.03$) \cite{dong} where
the only substantially 3D piece of the Fermi surface disappears ($x\sim
0.05).$ At the same doping level, magnetic ordering disappears, very much in
line with standard intuition that long range order in a 2D system is
easily destroyed by fluctuations. Superconductivity appears at the same time, suggesting, rather
naturally, that at smaller $x$ it is simply suppressed by the magnetic
order. The nice correlation between dimensionality, magnetism, and
superconductivity is entirely destroyed in calculations based on the
experimental structure. The 3D band that crosses E$_{F}$ and forms a small
pocket around the Z point of the BZ (see Fig. \ref{oeband}) in the optimized
structure is far below E$_{F}$ in the experimental structure, replaced by a
strongly 2D band. In other words, expanding the Fe-As distance exchanges the
positions of a 3D and a 2D band, causing the latter to cross E$_{F}$. The
system is then two-dimensional for all values of $x$ and doping is expected
to have little effect on the strength of the spin fluctuations. Of course,
at some doping this 2D band will also fill (similar to the 3D band of the
optimized structure), but no dimensional crossover will occur coincident
with the disappearance of the Fermi surface.

Although the experimental trends can seemingly be better explained by using 
the GGA-optimized coordinates followed by LDA magnetic calculations than by 
using GGA magnetic calculations or experimental coordinates, as outlined 
above, the exception of the coordinates themselves is important. 
Surprisingly, applying spin polarization {\it and} the GGA 
exchange-correlation potential during optimization leads to rather good 
agreement with experiment for internal positions as well as lattice constants 
(the latter are already reasonably reproduced in paramagnetic GGA 
calculations).  The Fe-As height, to which the FS and magnetism are unusually 
sensitive, is obtained as 2.38 $\AA$, acceptably close to the experimentally observed 
height of 2.41 $\AA$.  Thus, there is a contradiction between good agreement in 
structural parameters and wildly overestimated magnetism (or conversely, good 
magnetic energies/moments and badly underestimated bond-lengths) that remains 
one of the more confusing aspects of this compound.  In order to fully 
understand how magnetism manifests itself in this family of materials, which 
is crucial as a beginning step to understand the relationship between 
magnetism and superconductiviy, the connection between the physical and 
magnetic structures must be disentangled.  This remains an open and 
intriguing problem.

We now address the origin of the antiferromagnetic interactions. As pointed
out in Ref. \cite{mazin}, all three magnetic instabilities have entirely
different physical origins: ferromagnetism is of the Stoner type,
checkerboard antiferromagnetism results from the combined effort of
conventional superexchange and of nesting of two electron pockets\cite{nest}, while the
stripe ordering appears because of nesting between the hole and electron
pockets, and is additionally supported \cite{yildirim} by the next-n.n.
superexchange. It has been assumed by some authors \cite%
{yildirim,ma2,si,kivelson} that superexchange is the leading cause of both
AFM instabilities, and, by implication, that a two-shell Heisenberg model
provides a meaningful description of magnetism. Others \cite
{mazin,dong,kuroki} emphasized the long-range character of magnetic
interaction, coming from interband spin susceptibility. The physical
difference between the two descriptions is enormous: the former implies AFM
in real space, that is, each two individual ions that are neigbors or next
neighbors to each other interact antiferromagnetically, but there is no
direct interaction at longer distances. The latter interaction is local in
the reciprocal space, that is, it corresponds to condensation of a
spin-density wave (even though only a $finite$ amplitude SDW is stable), and
requires ordering on the scale of at least several unit cells to exhibit
antiferromagnetism. Direct evidence in favor of the long-range SDW scenario
comes from the calculations of Yin {\it et. al} \cite{pickett}: they have
computed nn and nnn exchange constant in the ordered stripe phase \textit{%
with respect to the small deviations from the AFM ordering.} They found that
the exchange constants have opposite signs along the CDW wave vector and in
the perpendicular direction. Given that the underlying structure is
tetragonal and the x/y asymmetry arises entirely due to the magnetic
ordering itself, this proves beyond any doubt that the Heisenberg
Hamiltonian is not applicable for this system.

We have also undertaken a more conventional test that also confirms the SDW
character of the stripe antiferromagnetism. First, one can try to map the
magnetic energy of the two stable AFM structures onto the n.n. Heisenberg
model. This is a routine procedure for localized systems. Here, however, it
has the problem that the moments are so soft that not all desired magnetic
patterns can be realized, most notably the FM structure, and different
patterns converge to different magnetic moment amplitudes (extreme softness
of the magnetic moment magnitude 
already implies that the application
of the Heisenberg Hamiltonian is suspect). This problem was
realized by Yildirim who added an external field (when needed, staggered) to
converge all three patterns he considered at the same moment magnitudes. The
hope is that the result is not too sensitive to the shape of the staggered
field (which is true for localized moments, but not necessarily for
itinerant ones). Ma, on the other hand, allowed the moments to vary freely
within each of the three antiferromagnetic orderings, and found magnetic
moments ranging from 2.2 - 2.6 $\mu _{B}$, and then adjusted the
ferromagnetic structure to have approximately the same moment.

In the checkerboard structure, the Heisenberg (or Ising, in this case)
exchange energy per site, $\sum_{i>j}J_{ij}M_{i}M_{j},$ is $
E_{cb}=-2(J_{nn}-J_{nnn})M^{2},$ and in the stripe phase $
E_{st}=-2J_{nnn}M^{2}$ (with respect to the nonmagnetic state). It is also
possible\cite{ma2} to stabilize an intermediate magnetic structure, where
the sites with the coordinates ($2n,2m)$ have spins up, $(2n+1,2m+1)$ spins
down, and $(2n+1,2)$ and $(2n,2m+1)$ are not magnetic. In that case the
magnetic energy per site is half that of the stripe phase, $-J_{nnn}M^{2}$.
Finally, the ferromagnetic state energy in the Heisenberg model is $
+2(J_{nn}+J_{nnn})M^{2}.$ Ma $et$ $al$ \cite{ma2} give for the
ferromagnetic, checkerboard, stripe and the intermediate structures the
energies of $+91,$ $-217,$ $-109$ and $-65$ meV/Fe, respectively. It is
obvious that these four numbers are nowhere close to be mappable onto this
model, contrary to the claim in Ref. \cite{ma2}. It is also worth noting
that the very fact that the checkerboard structure is stable in GGA implies $
J_{nn}>J_{nnn},$ contrary to the popular belief\cite{si,kivelson} that the nnn superexchange
is stronger here.

One may notice that the energies above are inconsistent with any published
all-electron calculation, so one may agree with Yildirim\cite{yildirim} that
in this case pseudopotential calculations cannot be trusted (although our own
experience indicates that \textit{carefully and properly selected }
pseudopotentials are still reasonable). To test that, we have performed
similar calculations using the all-electron WIEN code, also using the
optimized structure, $x=0$ and GGA (which makes sense in this particular
case in order to artificially emphasize the magnetism). We found the four
energies in question to be $+166$, $-18,$ $-81$ meV and $-36$ meV (and the
moments ranging from 1.5 to 1.8 $\mu _{B}).$ These numbers are reliable DFT
results, with no uncontrollable approximations such as pseudization of the
potentials, yet they are just as incompatible with the two-neighbor Heisenberg
model as are Ma $et$ $al$'s results, in agreement with the perturbative calculations
of Yin $et.$ $al$.

We conclude this section by restating the main points relevant to magnetic calculations in this system. In both GGA and LDA the 
undoped system is on the verge (within a fraction of a meV) of an instability against an itinerant ferromagnetic state with a very 
small magnetic moment. The AFM checkerboard structure, reminiscent of the magnetic ordering of Cu planes in cuprates, is quite 
stable, albeit not the ground state, in both LDA and GGA, if one uses experimental As positions, in either the doped or undoped 
compounds. In the optimized structure the checkerboard magnetism is stable only in GGA, and by a rather small energy. The striped 
structure is the DFT ground state for all relevant dopings, independent of the As position; in GGA it is very stable, while in the 
LDA, and if one uses optimized positions, it rapidly becomes a borderline instability that can be easily destroyed by fluctuations, 
in agreement with the experiment. Magnetism cannot be described, not even qualitatively, using two superexchange coupling constants, 
not even if one adds a Stoner energy reflecting the softness of individual spins. Only LDA calculations, and only in the optimized 
structure, provide a quantitatively realistic picture of the magnetic properties of this system in the entire doping range. It is 
worth reiterating that in borderline magnets, of which this system is clearly an example, not only is LDA closer to experiment than 
GGA, but in real materials fluctuations suppress magnetism even further, so that in comparison to GGA, LDA is a step in the right 
direction, but does not go the whole way.

\begin{figure}[tbp]
\includegraphics[width= 0.95 \linewidth]{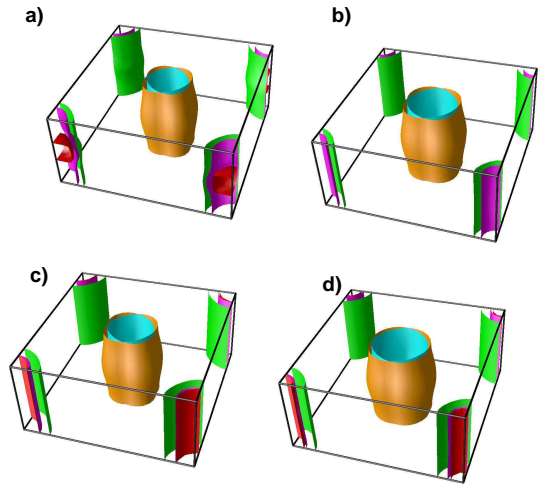}
\caption{(color online) GGA Fermi surfaces of LaFeAsO$_{1-x}$F$_x$, for a) $x$=0 with optimized coordinates, b) $x$=0.1 
with optimized coordinates, c) $x$=0 with
 experimental coordinates, and d) $x$=0.1 with experimental coordinates.  Note that
the 
 band structure is 
three dimensional at low doping
 in the optimized structure, but  becomes two-dimensional with doping,
while
in the experimental structure it is two-dimensional
 and insensitive to doping}
\label{fs-fig}
\end{figure}

In the second part of this work, we concentrate on a detail of the Fermi
surface that is crucially important for some of the proposed
superconductivity theories, as well as for some of the magnetic orderings:
 the eccentricity of the overlapping ellipses
around the $M$ point of the BZ. The importance of the eccentricity is
obvious from the fact that if the electron pockets were ideal circles
(though they do not have to be, by symmetry), just as the hole pockets, they
would nest ideally with each other, creating a sharp nesting peak at $Q=\pi
,\pi $ and strongly enhancing the tendency to checkerboard AFM ordering. At
the particular doping where the number of holes is exactly equal to the
number of electrons, the hole and the electron pockets would also nest
ideally, making this specific concentration very special from the electronic
point of view. In addition, such sharp structure in the magnetic
susceptibility is favorable for exotic order parameter distributions with
nodes, set not by symmetry but by resonance effects between the peaks in the
susceptibility and the Fermi surface geometry \cite{kuroki}. At least one of
the proposed theories\cite{dai,zhang} relies on the condition that two
rotated electron pockets coincide with an accuracy in energy not worse than
the superconducting gap.

Although the band structure and fermiology are much less sensitive to
calculational method than the stabilization energies are, there is still
some variation between methodologies and substantial differences occur
between published reports (in the latter case, largely due to variance
among the structures used by different authors).

\begin{table}[tb]
\caption{Calculated 
 eccentricities, $e(\Gamma )$ and $e(Z)$ of the
electron Fermi surfaces as a function of doping level with different
methods. Here 
 $e(\Gamma )$ and $e(Z)$ are in the $k_{z}$=0 and $k_{z}$=0.5 planes,
respectively. Note that at low doping there is an additional 3D hole sheet.}
\label{table_ecc}\vspace{0.2cm} 
\begin{tabular}{llccccc}
\colrule &  & ~~$e(\Gamma)$~~ & ~~$%
e(Z)$~~ \\ \hline
FP & $x$=0 & 0.34 & 0.59 \\ 
(Wien) & $x$=0.1 & 0.26 & 0.48 \\ \hline
FP & $x$=0 & 0.31 & 0.61 \\ 
(FPLO) & $x$=0.1   & 0.23 & 0.48 \\ \hline
PAW & $x$=0   & 0.28 & 0.52 \\ 
(VASP) & $x$=0.1   & 0.18 & 0.42 \\ \hline
PP & $x$=0   & 0.20 & 0.48 \\ 
(PWSCF) & $x$=0.1   & 0.15 & 0.42 \\ \hline
PP \cite{kuroki} & $x$=0   & 0.20 &  \\ \hline
PP \cite{zhang} & $x$=0   & 0.09 & 0.42 \\ 
& $x$ = 0.1   & 0.0 &  \\ \hline
PP \cite{cao} & $x$=0   & 0.26 &  \\ \hline
\end{tabular}%
\label{tab-eccentricity}
\end{table}

We have summarized our own results, again using two all electron
full-potential codes and two pseudopotential codes in Table \ref%
{tab-eccentricity}, where we additionally include eccentricities extracted
from other reported calculations. The eccentricity in the $\Gamma -X-M$ ($%
k_{z}$ = 0) plane is listed, and, where available, in the $Z-R-A$ plane ($%
k_{z}$ = 0.5). The latter is universally larger, due to the fact that the
longer axis of the ellipse is formed by the only band around the M point that is
(somewhat) $z$-dispersive.

Again we see excellent (nearly perfect) agreement between our two all
electron results with more deviation in the pseudopotential ones. In fact,
all pseudotential calculations appear to underestimate the eccentricity,
some by a large amount. Ref. \onlinecite{zhang} has reported zero
eccentricity at a doping of $\sim 10\% $ in the $\Gamma$ plane, far less
than not only both all-electron calculations, but also other pseudopotential
results. Surprisingly, in the $Z$ plane, the eccentricity very closely
matches our pseudopotential results, indicating a highly dispersive band
absent from our calculations. This is in some sense unfortunate, because a
very interesting model of interband $triplet$ s-wave pairing \cite{dai} is
only viable if the energy difference between the two bands at the M point is
smaller near the point where they cross the Fermi level, than the
superconducting gap. Our all-electron results unequivocally indicate that
this difference is at least 100 meV in the $k_{z}=0$ plane and at least 180
meV in the $k_{z}=\pi /c$ plane.

Several things should be noted when comparing these calculations. The first
is that doping is accounted for differently in our pseudopotential
calculations than in our all-electron calculations. In the former, the
desired number of electrons are added and compensated for with a uniform
positive background. In the latter, electron doping is obtained using a
fictitious O atom with increased core charge and electrons and hole doping
is obtained using a fictitious La atoms with decreased core charge and
electrons. Thus, one should not expect the two different methodologies to
produce identical results. The second thing is that for calculations other
than our own in Table \ref{table_ecc}, different lattice constants or
internal coordinates may have been employed, although it is our experience
that, within reasonable limits, these never drastically reduce the
eccentricities (even though the actual numbers change).

In conclusion, we show that while the magnetic ground state of LaFeAsO$%
_{1-x} $F$_{x}$ can be obtained using any of several methodologies
(all-electron, pseudopotential with ultra-soft or pseudopotential with PAW),
the details of the magnetic properties, such as site magnetization and
magnetic energy differences may be affected by pseudization of the crystal
potential. More importantly, the results depend drastically on the
exchange-correlation functional used (GGA vs. LDA) and on the position of
As. The best results in terms of explaining the observed magnetic phase
diagram are obtained within LDA (which is the recommended functional for
itinerant magnets) and using the theoretical As positions. The fermiology
and band structure are less sensitive to the details of the calculation,
with the three all-electron codes we applied giving identical results. Small
Fermi surface details, such as the eccentricity of the central ellipses,
 differed between all-electron and pseudopotential codes
and between different pseudopotential codes, though a careful application of
the latter yielded results that differ only slightly from the all-electron
results. We have reported a reference set of calculations for the basic
electronic structure properties of the parent compound and for several
dopings (both hole and electron). We believe these will provide a
consistent and accurate basis of knowledge upon which models and theories
about this interesting compound can be built.

\acknowledgments

Work at NRL was supported by the Office of Naval Research. Work at ORNL was
supported by the Department of Energy, BES, Division of Materials Sciences
and Engineering.

\

{\it Note added in proof:}  After submission of this manuscript, we became 
aware of a paper by Ishibashi {\it et. al} \cite{ishibashi}.  These authors 
applied an in-house pseudopotential code to four different fully AFM patterns 
($\mu \geq$ 1.8 $\mu_B$ on all sites).  Their stabilization energies in all 
four cases agree with our all-electron results within 7 meV/formula unit.  
Curiously, if one discards the FM pattern, the three energy differences among 
the remaining AFM states can be fitted with a two-parameter Heisenberg model 
fairly well.  However, the fact that the FM ordering does not fit into this 
picture at all suggests that the fit is fortuitous.


\begin{thebibliography}{99}
\bibitem{nest} Here and below we use the word ``nesting'' in a loose 
way, applying it to any interaction derived from lowering energy by
removing one-electron states from the Fermi level. It is important to
realize that the peaks in the calculated one-electron susceptibility are not sharp
enough to induce instability with respect to an infinitesimaly small
perturbation with the ``nesting'' wave vector, whether for the stripes
of for the checkerboard structure. The instability occurs only for SDWs of large
amplitudes, in which case the perfect nesting exactly at the Fermi level
is less important.
\end{thebibliography}
\end{document}